\documentclass[aps,prl,twocolumn,amsfonts,showpacs,superscriptaddress]{revtex4} 
\usepackage{bm}
\usepackage{epsfig,amsopn}
\usepackage{graphicx}
\usepackage{amsmath,amssymb}
\usepackage{natbib}
\renewcommand{\section}[1]{{\par\it #1.---}}
\def\be{\begin{equation}}
\def\ee{\end{equation}}
\def\bea{\begin{eqnarray}}
\def\eea{\end{eqnarray}}
\def\la{\langle}
\def\ra{\rangle}
\def\om{\omega}
\def\nn{\nonumber}
 
\def\cH{{\mathcal{H}}}
\def\cI{{\mathcal{I}}}
\def\f{\frac}
\def\g{\gamma}
\def\al{\alpha}
\def\etal{et al.}
\begin{document}

\title{Fluctuation theorem in quantum heat conduction}
\author{Keiji Saito}
\affiliation{Graduate School of Science, 
University of Tokyo, 113-0033, Japan} 
\affiliation{CREST, Japan Science and Technology (JST), Saitama, 332-0012, Japan}
\author{Abhishek Dhar}
\affiliation{Raman Research Institute, Bangalore 560080, India}
\date{\today}

\begin{abstract}
We consider steady state heat conduction across a quantum harmonic
chain connected to reservoirs modelled by infinite collection of
oscillators. The heat, $Q$, flowing across the oscillator in a  
time interval $\tau$ is a stochastic variable and we study the 
probability distribution function $P(Q)$. 
We compute the exact generating function of $Q$ at large $\tau$ 
and the large deviation function. The generating function
has a symmetry satisfying 
the steady state fluctuation theorem without any quantum corrections.
The distribution $P(Q)$ is nongaussian with 
clear exponential tails.
The effect of finite $\tau$ and nonlinearity is considered in the
classical limit through Langevin simulations. 
We also obtain the prediction of quantum heat current fluctuations at low temperatures in clean wires.
\end{abstract}
\pacs{05.40.-a,05.70.Ln,44.10.+i,63.22.+m}
% 02.50.-r Probability theory, stochastic processes, and statistics 
% 05.30.-d Quantum statistical mechanics  
% 05.40.-a Fluctuation phenomena, random processes, noise, 
%          and Brownian motion(1)
% 05.60.Gg Quantum transport  
% 05.70.Ln Nonequilibrium and irreversible thermodynamics (2)
% 65.40.Gr Entropy and other thermodynamical quantities  
% 67.40.Pm Transport processes, second and other sounds, 
%       and thermal counterflow; Kapitza resistance  
% 63.22.+m Phonons or vibrational states in low-dimensional 
%          structures and nanoscale materials (4)
% 44.10.+i Heat conduction (3)
% 63.20.-e Phonons in crystal lattices 
%%

\maketitle 
%\begin{itemize}
%\item mention on quantum correction
%\item why may quantum correction be needed?
%\item Jartynski equation, TFT are both needed, then SSFT ?
%\item Monnai-Tasaki, quantum case are still controversial.
%\item Do not mention about electron transport much.
%\item Against referee A, how should we make introduction physical?
%\item Stress that fundamental nonequilibrium process is common to all fields.
%\end{itemize}

A lot of interest has been generated recently in
fluctuations in entropy production in nonequilibrium systems.
Several definitions  of entropy production have been used and these
give some measure of ``second law 
violations''.
A number of authors
have looked, 
both theoretically \cite{evan1,lebo99,jarz97,zon04} 
and in experiments \cite{wang,JGC06}, at fluctuations of quantities
such as work, power flux, heat absorbed, etc. during nonequilibrium
processes and these have been generically referred to as entropy
production. The new results, referred to as the fluctuation theorems,
make general predictions on the probability 
distribution $P(S)$ of the  entropy $S$  produced during a
nonequilibrium process  
\cite{evan1,lebo99}. 
Specifically these theorems quantify the probability of negative entropy
producing events which become 
significant if one is looking at small systems or at small time
intervals. There are two different 
theorems, the transient fluctuation theorem (TFT) and 
the steady state fluctuation theorem (SSFT).
The TFT looks at the entropy produced in a finite time 
$\tau$ in a non-steady state.
In SSFT, one looks at 
a nonequilibrium steady state and the average entropy
production rate over a long time interval $\tau$ is examined. 
The precise statement of SSFT is:
\bea
\lim_{\tau \to \infty} {1 \over \tau} \ln \Bigl[ 
\f{P(S= \sigma \tau)}{P(S=-\sigma \tau)}\Bigr] =
{\sigma }~. \label{SSFT}
\eea
In the context of SSFT a quantity of great interest is the large
deviation function ${h}(\sigma)$ which specifies the asymptotic form of
the distribution function $P(S)$ through the relation $P(S)\sim
e^{\tau h(\sigma) }$ \cite{lebo99,derrida98}. An equivalent
statement of SSFT can be made in terms of a special
symmetry of $h(\sigma)$ which is:
$h(\sigma)-h(-\sigma)=\sigma$. Remarkably, this 
relation has been shown to lead to linear response results such as
Onsager reciprocity and the Green-Kubo relations \cite{GG96,lebo99,AG07}.  
Furthermore it leads to predictions for properties 
in the far from equilibrium regime. 

Heat conduction is a natural example where one talks of 
entropy production. The standard result
from nonequilibrium thermodynamics is that when an amount of heat $Q$
is transferred from a bath at temperature $T_L$ to a bath at
temperature $T_R ~(< T_L)$ the entropy produced $S$ is given by
$S=(T_{R}^{-1}-T_{L}^{-1})~Q$. However, in general $S$ is a stochastic 
variable with a distribution $P(S)$.
The distribution $P(S)$ for a nonlinear chain connected to
Nose-Hoover baths was studied numerically in \cite{lepri98} where they
verified that it satisfied SSFT.
Refn.~\cite{bellet02} studied heat conduction in a nonlinear 
chain connected to free phonon reservoirs.
Based on strong ergodicity properties of the model, it was 
proved 
that  $P(S) \sim e^{ \tau h(\sigma)}$ where 
$h(\sigma)$ satisfied the SSFT symmetry. In a set-up with direct 
tunneling between two finite systems, a
transient version of the heat exchange fluctuation theorem, valid both
for classical and quantum systems, was proved in \cite{jarz04}. 

While the SSFT clearly presents a powerful theorem for nonequilibrium
systems, its validity has been established only in specific systems
and so far only classically. 
%AD: Drop this statement? Repeated after next line. 
% It is not clear whether quantum systems need quantum corrections or not. 
In the transient version,   
it was proved that quantum corrections are necessary for a  
 dragged Brownian particle \cite{monnai_tasaki}.
Thus it is an open question as to whether quantum corrections to  SSFT
exist in quantum heat transport and 
what the  characteristics of the heat current distribution are.
This letter presents the first explicit
calculation of $h(\sigma)$ and demonstration of SSFT in quantum heat 
conduction.
We study steady state of a quantum harmonic chain
connected to baths which are modelled by infinite oscillator sets.
This model is relevant to recent experiments 
on mesoscopic quantum heat transport\cite{angel98,schwab}, where
the quantized thermal conductance $g_{0} (T) =\pi k_{\rm B}^2 T /(6 \hbar )$ 
was measured\cite{rego98,schwab}. 
We use the method of full-counting statistics 
\cite{levitov} 
to compute the generating function of $Q$.
We then 
show that the corresponding large deviation function satisfies the
SSFT symmetry condition. 
For finite $\tau$ we consider heat
transport across small chains  and study the classical limit
through Langevin simulations. We also consider the effect of
introducing nonlinearity in the oscillator potential.  

\section{Model} Our model consists of a  harmonic chain coupled to
two heat baths kept at temperatures $T_L$ and $T_R$ respectively. For
the heat baths we assume the standard model of an infinite collection
of oscillators. The full Hamiltonian is given by
\bea
{\cal H} &=& \sum_{n=1}^N \left({p_n^2 \over 2m_n} 
+ { k_n \over 2}x_n^2 \right) +
\sum_{n=2}^{N}
  {k  \over 2} (x_n-x_{n-1})^2\nn \\ 
&+& \sum_{\ell} \left[ {p_{\ell}^{2} \over 2 m_{\ell}} +
{m_{\ell} \omega_{\ell}^{2} \over 2} 
\left( x_{\ell } - {\lambda_{\ell} x_1 \over m_{\ell} \omega_{\ell}^{2} 
} \right)^{2} 
\right] \nn \\
&+& \sum_{r} 
\left[
{p_{r}^{2} \over 2 m_{r}}
+ {m_{r} \omega_{r}^{2} \over 2} 
\left( x_{r } - {\lambda_{r} x_N \over m_{r} \omega_{r}^{2} 
} \right)^{2} 
\right]
, \label{original_hamiltonian}
\eea
where $\{m_n,x_n,p_n,k_n,k\}$
refer to the system degrees of freedom,
$\{x_\ell,p_\ell,m_\ell,\omega_\ell\}$ refers to the left reservoir while  
 $\{x_r,p_r,m_r,\omega_r\}$ refers to the right reservoir. The
coupling constants between the system and the bath oscillators
$\{\lambda_\ell,\lambda_r\}$ is switched on at time $t=-\infty$. The
initial density matrix is assumed to be of the product form
$\rho(-\infty)=\rho_S   \otimes \rho_L \otimes \rho_R$,
where $S,L,R$ refer respectively to the system and left and right
reservoirs. The left and right density
matrices are equilibrium distributions corresponding to the 
respective temperatures:
$\rho_\al={e^{-\beta_\al \cH_\al}}/{{\rm Tr} [e^{-\beta_\al
    \cH_\al}]}$ for $\al=L,R$ and $\beta_{\al}=1/(k_{\rm B}
T_{\al})$. 

It can be shown \cite{segal03} that eliminating the bath
degrees of freedom leads to 
an effective quantum Langevin equation  for the system.
The effect of the baths is to produce noise, given by $\eta_{L,R}(t)$,
and dissipative effects controlled by memory kernels $\g_{L,R}(t)$.
The properties of the noise and dissipation are completely determined
by the initial condition of the baths at $t=-\infty$. We now make a
few definitions. Let 
${\eta}_{L,R}(\om)=\int_{-\infty}^{\infty} dt
\gamma_{L,R}(t) e^{i \om t}$ ,
$\tilde{\gamma}_{L,R}(\om)=\int_0^\infty dt
\gamma_{L,R}(t) e^{i \om t}$ 
and let $\Sigma^r_{L,R}(\om)=-i \om
\tilde{\gamma}_{L,R}(\om)$, which, as we will see later, gives the
self energy correction coming from the baths to the Green's
function of the harmonic chain. 
We also define the spectral function
$J_{L}(\om)={\pi \over 2} \sum_\ell {\lambda_\ell^2 \over m_\ell
  \omega_\ell}\delta (\omega-\omega_\ell) 
$
for the left reservoir and a similar function $J_R(\om)$ for the right
reservoir. 
Then the dissipation kernels and noise correlations are given by:
\bea
\g_{\alpha}(t)&=&{2 \over \pi} \int_0^\infty d \om {J_\alpha (\om) \over \om}
\cos{\om t} \nn \\
\la \eta_\al(\om) \eta_\al(\om') \ra &=& 4 \pi \hbar \delta(\om +\om')
\Gamma_\al(\om)\,(1+f_\al(\om)) 
\eea
for $\al=L,R$ and where $\Gamma_{\al} (\omega ) =
-{\rm Im} \{\Sigma^r_{\al}(\om)\} =J_\al (\om)
 \Theta(\om) \nn - J_\al (-\om)
\Theta(-\om) $ and   $f_\al(\om)=1/(e^{\beta_{\al} \hbar \om}-1)$.
All higher noise correlations can be obtained from the two-point
correlator. Using the quantum Langevin approach it is
straightforward to derive the Landauer type result for  
average heat current $\la \hat{\cI}\ra$ \cite{segal03}: 
\bea
\la \hat{\cI} \ra &=& \f{1}{4\pi} \int_{-\infty}^\infty d \om {\hbar \om} 
{\cal T}(\om) 
      [~f_L(\om)-f_R(\om)~]~, \label{cur}\\
&&{\cal{T}}(\om) = 4 \Gamma_L(\om) \Gamma_R (\om)
      |G^r_{1,N}(\om)|^2 ~ ,\nn ~~~~\\
&&G^r(\om)=[{\bm M} \om^2 - {\bm K} -{\bm \Sigma}^r_L (\om )-{\bf
    \Sigma}^r_R(\om ) ]^{-1} ~,~~~~  
\eea
where ${\bm M}$, ${\bm K}$ are the mass and the force 
constant matrix and ${\bm  \Sigma}^r_{L,R}(\om)$ are 
self-energy correction matrices with
elements $[{\bm \Sigma}^r_L (\om ) ]_{m,n}=\Sigma^r_L (\om ) \delta_{m,n}
\delta_{m,1}$ and $[{\bm \Sigma}^r_R (\om ) ]_{m,n}=\Sigma^r_R (\om ) 
\delta_{m,n}
\delta_{m,N}$. Note that ${\cal {T}}(\om ) $ is the transmission coefficient
for phonons while $G^r (\om ) $ is the  phonon Green's function for the chain. 

\section{Statistics of phononic heat transfer}
The heat transfer in time $\tau$ is given by $\la \hat{\cI} \ra \tau$. 
Here we are interested in the statistics of heat transfer in the
nonequilibrium steady state and so need to calculate higher moments of
the heat transfer. 
%For this it is necessary to calculate higher order
%correlations such as $\la \hat{\cI}(t_1) \hat{\cI}(t_2) \ra$, etc. 
The quantum Langevin approach can in principle be used to compute correlation
functions at any order but this becomes increasingly
cumbersome. Instead we use the Keldysh approach which, as we will
show, gives the generating function of the heat transfer.

Several definitions of $\hat{\cI}$ are possible depending on where we
evaluate the current. Here we consider the current from the left
reservoir into the system ( 
obtained by taking a time-derivative 
of energy in the left reservoir $\cH_L=\sum_\ell [{p_\ell^2}/(2m_\ell)+m_\ell
\om_\ell^2 x_\ell^2/2]$):
\bea
\hat{\cI}=-\sum_\ell \f{\lambda_\ell}{m_\ell} p_{\ell} x_1 ~.
\eea
We also define the average heat transfer operator
$\hat{\cal{Q}}= \int_{-\tau/2}^{\tau/2} dt \hat{\cI}(t)$.
Using the Keldysh approach let us compute the following quantity:
\bea
Z(\xi)=
\Bigl\langle  \stackrel \rightarrow {\textsf{T}}
e^{
\left[ 
{i\over \hbar} \int_{-\infty}^{\infty} dt 
\left( {\cal H} - \varphi (t) {\cal I}\right)
\right]}  \stackrel \leftarrow {\textsf{T}} 
e^{\left[ 
-{i\over \hbar} \int_{-\infty}^{\infty} dt 
\left( {\cal H} + \varphi (t) {\cal I}\right)
\right]}
\Bigr\rangle~, ~ \label{formal_1} 
\nonumber 
\eea
where $\langle ... \rangle$ denotes an average over the initial state,
$\stackrel \rightarrow {\textsf{T}}$ 
and $\stackrel \leftarrow {\textsf{T}}$ 
denote forward and reverse time ordering, 
and the counting field $\varphi (t)$ is defined as
$\varphi (t) =
-{\hbar \xi / 2} ~~ {\rm for}~~ -{\tau/ 2} \le t \le {\tau/ 2}
$
and zero elsewhere.
It can be shown that $\ln{Z(\xi)}$ is the cumulant generating function 
for the heat operator:
\bea
\ln{Z(\xi)}= \sum_{n=1}^{\infty} {(i\xi )^{n} \over n! }
\langle \hat{\cal Q}^{n} \rangle_{c }~,
\eea
where $\la \hat{\cal Q}^n \ra_{c } $ is the $n^{\rm th}$ order cumulant
at large $\tau$. Hence
the probability distribution of measuring a heat transfer $Q$ is
obtained by taking the Fourier transform
$
P(Q)=\f{1}{2 \pi} \int_{-\infty}^\infty ~ d\xi~Z(\xi)~e^{-i \xi Q}~. 
%\label{PQ}
$
For large $\tau$ one obtains $Z(\xi) \sim e^{ \tau {\cal G}(\xi) }$ and $P(Q)
\sim e^{\tau  \tilde{h}(q)}$ with $\tilde{h}(q)=   {\cal G}(\xi^*)
-i\xi^* q$ and where $\xi^*$ is the solution of the saddle-point equation 
${d {\cal G}(\xi^*) / d \xi^*} -i q = 0$. 
We evaluate ${\cal G} (\xi)$ using standard path integral and Green's
function techniques along the Keldysh contour. 
The final result is the following form:
\bea
 {\cal G}(\xi ) &=&
{-1 \over 4 \pi} \int_{-\infty}^{\infty} 
\hspace*{-0.25cm} d\omega \ln 
\Bigl\{  1 + {\cal T}(\om) [ 
 f_{R}(-\om) f_{L}(\om) ( e^{i \xi \hbar \omega }  - 1 ) 
 \nn \\
&+& 
 f_{ R}(\om) f_{ L}(-\om)
( e^{-i \xi \hbar \omega }  - 1)  
]  
 \Bigr\}~. \label{ldf}
\eea
Phonons convey energy in units $\hbar\omega$ and this appears in the
exponential form with the factor $\xi$.
It is easily verified that Eq.~(\ref{ldf}) reproduces the correct
first moment of 
$\hat{\cI}$ given in Eq.~(\ref{cur}). The second moment is given by
\bea
\f{\la \hat{\cal Q}^2 \ra_c}{\tau}&=&\f{1}{4\pi} \int_{-\infty}^\infty
d \omega ~(\hbar \om)^2 
\Bigl\{ {\cal T}^2 (\omega ) 
~[ f_L(\om)-f_R(\om) ]^2 
\nn \\ &-& 
{\cal T} (\omega ) 
[ f_L(\om) f_R(-\om)
  +f_L(-\om) f_R(\om) ] \Bigr\} . \label{mom2}
\eea
We have verified this also with the Langevin approach. 
This bosonic fluctuation is similar to the optical one \cite{buttiker00}.
\section{Symmetry}
We note the following symmetry of $\cal G$:
\bea
{\cal G}(\xi)= {\cal G}
\left(\, -\xi+i{\cal A}\,\right) ~, \label{symmofg}
\eea
where ${\cal A} =\beta_R  - \beta_L $. 
Using the identification $\sigma={\cal A}q$ and the relation between
$h(\sigma)=\tilde{h}(q)$ and ${\cal G}(\xi)$  immediately leads
to the SSFT relation Eq.~(\ref{SSFT}).
Thus we conclude that quantum heat transports satisfy the SSFT without 
any quantum corrections.

The symmetry (\ref{symmofg}) contains
information regarding transport coefficients \cite{AG07}. 
For fixed $\beta_L + \beta_R$ let us make the expansion $\la
{\hat{\cal{Q}}^n}\ra_{c}/\tau= \sum_m L_{n,m} {\cal A}^m / m!$. 
The nonlinear response coefficients $L_{n,m}$ are then given by
$L_{n,m} = {\partial^{n+m} {\cal G}(\xi ) / \partial (i\xi)^{n} 
\partial {\cal A}^{m}}\vline_{\,\xi={\cal A}=0}$. 
This coefficient represents a nonlinear response 
of the general cummulants of current to the 
thermodynamic force $(\beta_R - \beta_L)$.
The symmetry (\ref{symmofg}) gives the general
relations between the coefficients: 
\bea
L_{n,m} &=&
\sum_{k=0}^{m}\,\left(
\begin{array}{c}
m \\k
\end{array}\right)
(-1)^{(n+k)}  
 L_{n+k,m-k} , ~~~~~ \label{second}
\eea
with $L_{0,m}=0$.
For example we get $ L_{2,0}= 2 L_{1,1}$ and 
$L_{4,0}=2 L_{3,1} =6L_{2,2}-4L_{1,3}$. The first relation is simply
the Green-Kubo formula relating the linear current response to
equilibrium fluctuations while the second 
leads to relations between nonlinear response coefficients.

\begin{figure}
\includegraphics[width=3in]{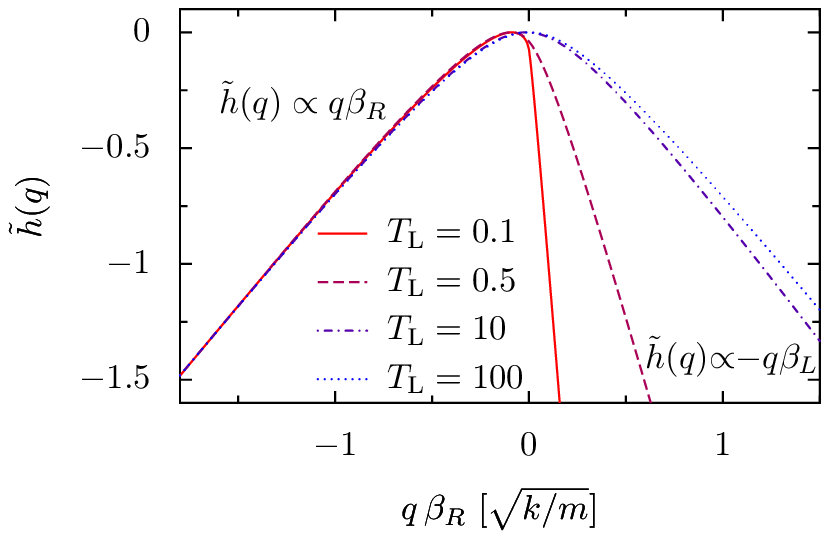}
\caption{Plot of $\tilde{h}(q)$ for various temperature regime
with $T_{R}-T_{L}$ fixed to 1.0$~[\hbar k_{\rm B}^{-1}\sqrt{k/m}]$. 
The parameters:$N=2, m_{1}=m_{2}=m, k_1=k_2=k,$ and $\gamma=1.0~[\sqrt{mk}]
$.}   
\label{fig1}
\end{figure}
\section{Typical distributions}
We present some results on the form of the distribution $P(Q)\sim
e^{\tilde{h}(q)}$  
for a small chain ($N=2$) connected to ohmic
reservoirs ($\tilde{\gamma}_{L,R}(\om)=\gamma$).
In Fig.~\ref{fig1} we plot $\tilde{h}(q)$, which is numerically obtained, for 
different temperatures $T_L$ with fixed temperature difference $T_{R}-T_{ L}$. 
In all temperature regimes, $\tilde{h}(q)$ shows a clear linear dependence
at large $q$, and those are well fitted by $\beta_{R} q$ and 
$-\beta_{L}q$ for $q<0$ and $q>0$ respectively.
This exponential tail is one of the characteristics in $P(Q)$.

\begin{figure}
\includegraphics[width=3.3in]{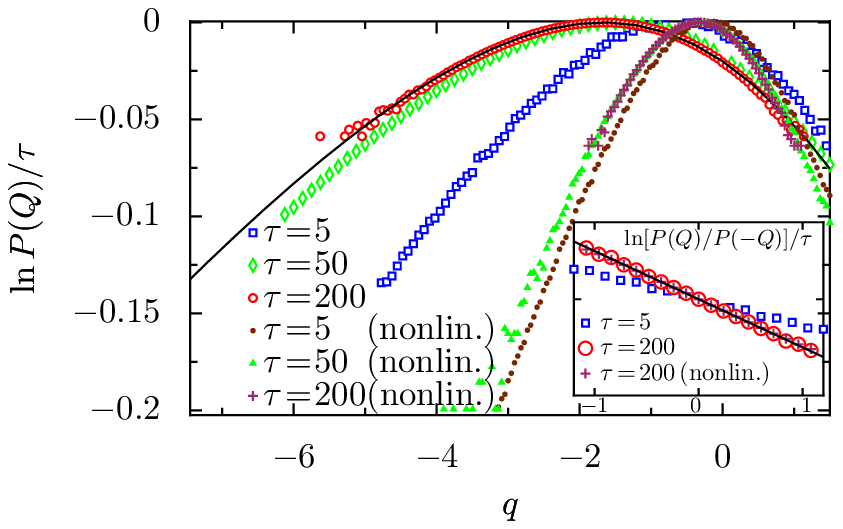}
\caption{Plot of $\ln[P(Q)]/\tau$ from Langevin simulations for the same
   system as  in Fig.~\ref{fig1} for various $\tau$.
Here $P(Q)$ is normalized so that its maximum is one. The large
   deviation function $\tilde{h}(q)$ is also shown by black solid line. 
The parameters are $N=2,m=k=1$, $k_{\rm B} T_L=10$, and $k_{\rm B} T_R=20$. 
The nonlinear case  has $\alpha=2$.
The inset shows $\ln [ P(Q) /P(-Q) ]/\tau$, and 
SSFT line $q(\beta_R - \beta_L)$ (black solid line).
Heat transfer is measured at the contact to the left reservoir. 
}   
\label{fig2}
\end{figure}

We now study  the effects of a finite $\tau$ and nonlinear potential
using the classical system. We evaluate $P(Q)$ from direct
simulations of the classical Langevin equations with white noise. 
In Fig.~\ref{fig2} we compare the
simulation results for different values of $\tau$ with the asymptotic
function $\tilde{h}(q)$ (obtained for $\hbar\to 0$).
It is clear from Fig.~\ref{fig2} that convergence
to the asymptotic distribution function takes place on a
rather large time scale. The nonlinear case is also
plotted for the same system with an onsite potential 
$V(x_{n})=\alpha  x_{n}^{4}/4$. In the inset, the function $\ln
[ P(Q)/P(-Q)]/\tau$ is plotted for three cases. 
The distribution for the nonlinear case deviates from the harmonic
cases, and both the average  
heat current and its fluctuations  are suppressed. However, as the
inset shows, SSFT  is satisfied in the nonlinear case,
which indicates the symmetry (\ref{symmofg}) and the relation (\ref{second})
hold too. 

\section{Heat current fluctuations in a pure wire}
Using Eq.(\ref{ldf}), we can derive the heat current fluctuations   
for a homogeneous wire connected to reservoirs through non-reflecting
contacts, a case for which the quantized thermal conductance has been measured 
\cite{schwab}. 
Consider a pure wire with all masses and spring constants equal. 
If we consider that the heat reservoirs themselves are 
semi-infinite wires (\emph{i.e.}, the Rubin model of a heat bath)
then it is easy to show that the contacts are perfect and we get
${\cal{T}}(\om)=1$ for all $\om$ within the allowed bandwidth. 
At low temperatures and for small $\Delta 
T=T_L-T_R$, Eq.~(\ref{cur}) leads to the quantized heat conductance
$g_{0} (T) =\la \hat{\cI}\ra/\Delta T=\pi k_{\rm B}^2 T /(6 \hbar )$. 
From Eq.~(\ref{mom2}) 
we now also get the thermal noise power at
zero frequency $S_0 =\la\hat{\cal Q} ^2 \ra/\tau $:
\be 
S_0 = k_B T_L^2 g_0 (T_L ) + k_B T_R^2 g_0 (T_R ) .
\ee
This is valid for $T_{L,R}$ in the temperature regime where
$g_0(T)$ can be measured \cite{schwab}. 
The noise power is also independent of details of system.
Independent contributions from $T_L$ and $T_R$ are obtained 
since there are no scattering process between phonons.  
Eq.~(\ref{ldf}) with ${\cal T}(\om )=1$ gives us the 
generating function valid in the same regime.
While our results have been derived for a one-dimensional wire with
scalar displacement variables they
are easy to generalize. Similar results can be obtained for realistic
models \cite{yama} of nanowires and nanotubes. 

\section{Summary}
Unlike equilibrium physics there are few general principles to
describe nonequilibrium phenomena. The exceptions to this  are 
the Onsager reciprocity  and the Green-Kubo relations which are valid
in the close-to-equilibrium linear response regime.  In view of this 
the nonequilibrium fluctuation theorems are quite remarkable in that
they seem to be exact relations valid arbitrarily far from equilibrium
and from which one can recover standard linear response theory. 
However the full range of validity and applicability of these theorems
is still not known.  
In this paper we have derived the explicit distribution for
fluctuations in phononic heat transfer across a quantum harmonic chain 
and have obtained the first proof of SSFT in quantum heat
conduction. We find that there are no quantum corrections.
%AD: Am dropping this sentence. Please insert if necessary.
%We have also made predictions of heat current fluctuations for a clean wire. 
We note that 
fluctuations in charge current in mesoscopic systems have been 
already studied both theoretically 
\cite{levitov,buttiker00,roche05,utsumi07} and experimentally \cite{expt}.
The present study provides a theoretical basis to study flucutuation of heat 
transfer. 
The measurement of fluctuations of phononic heat transfer in experiments
is an important challenging problem. 
A modification of the set up used in \cite{schwab} should be able
to make a measurement of fluctuations in heat transfer. 
One possibility is to use heaters with some
feed-back mechanism so as to maintain the two reservoirs at fixed
temperatures. The fluctuations in the power from the heater would be
related to the fluctuations in the heat transfer through the wire. 

\begin{acknowledgments}
We thank Y. Utsumi for useful comments. 
We thank IMS, National University of Singapore where this work was
initiated. AD thanks IMS for supporting a visit.
KS was supported by
 MEXT 
%the Grant in Aid from 
%the Ministry of Education, Sports, Culture and Technology of 
%Japan
, Grant Number (19740232).
\end{acknowledgments}

\end{document}